\documentclass[pdflatex,sn-mathphys-num]{sn-jnl}
\usepackage{graphicx}%
\usepackage{multirow}%
\usepackage{amsmath,amssymb,amsfonts}%
\usepackage{amsthm}%
\usepackage{mathrsfs}%
\usepackage[title]{appendix}%
\usepackage{xcolor}%
\usepackage{textcomp}%
\usepackage{manyfoot}%
\usepackage{booktabs}%
\usepackage{algorithm}%
\usepackage{algorithmicx}%
\usepackage{algpseudocode}%
\usepackage{listings}%
\usepackage{epstopdf}%

\DeclareUnicodeCharacter{03B8}{\ensuremath{\theta}}
\DeclareUnicodeCharacter{03C0}{\ensuremath{\pi}}

\begin{document} 
\title{Surface Variables Description of Axion Topological Materials } 

\author*[1]{\fnm{Francisco J.} \sur{Solis}}\email{francisco.solis@asu.edu}

\author[2]{\fnm{Hugo} \sur{Garc\'{\i}a-Compe\'an}}
\email{hugo.compean@cinvestav.mx}

\affil*[1]{School of Mathematical and Natural Sciences, Arizona State University, Glendale, Arizona,  85306}
\affil[2]{ Departamento de F\'{\i}sica, Centro de Investigaci\'on y de Estudios Avanzados del IPN,
 P.O. Box 14-740, CP. 07000, Ciudad de M\'exico, Mexico}








\abstract{
We study the response of axion topological materials under the presence of external static electric and magnetic fields. We focus on the macroscopic quasi-static magnetoelectric response of topological insulators. We use  techniques based on surface variables that have been previously employed in soft condensed matter problems for the description of heterogeneous systems composed of multiple homogeneous materials. A complete description of the whole system is  written in terms of surface degrees of freedom, which in this case correspond to effective surface charge and surface current densities. We obtain a set of integral equations satisfied by these variables. We present exact analytic solutions for axial-symmetric cases. We develop a numerical method for the solution of the integral equations by means of a boundary finite element method. We apply the numerical method to topological insulators with different geometries such as spheres, hollow spheres and toroidal surfaces. We show that that a  variational principle can be used to recover the surface variables equations.}

\maketitle
\newpage


\newpage

\section{Introduction}
\label{S-Intro}

Topological materials are substances in which quantum effects are manifested topologically. These have been extensively studied, as reviewed for example in refs.  \cite{hasan_colloquium_2010,Qi:2010qag,Yan:2016euz,witten_fermion_2016,witten_three_2016,planelles_axion_2021,sekine_axion_2021}. Examples of these  materials include topological insulators, topological superconductors, and topological Weyl semimetals. These systems can be described  by means of effective field theories that include axion electrodynamics. For topological insulators and superconductors, which are gap in the bulk and gapless in the boundary, the axion field is constant within the material \cite{wilczek_two_1987,witten_dyons_1979_alt,canfora_2011_alt,huerta_optical_2012_alt}. Topological Weyl semimetals, which are gapless in the bulk, also have an effective action with axion electrodynamics in which  the axion field is position and time dependent. The effective theories for these materials include terms in their Lagrangian that violate Lorentz symmetry \cite{colladay_cpt_1997,colladay_lorentz-violating_1998}. Surface charges and currents (electronic states) are real and persistent quantities due to topological reasons and they are measured in several topological insulator materials. In this work we focus on the case of topological insulators where the sharp changes in axion field define clear boundaries. In this case the axion field $\theta$ takes the value $\theta=\pi$ in the bulk of the insulator and is zero outside the material. We  show that knowledge of values of the surface variables is sufficient to reconstruct the properties of the whole system. 

The description of a topological insulator in terms of axion  electrodynamics can be studied through different methods. One of the most powerful approaches for the case of an arbitrary configuration of sources is the Green function method \cite{martin-ruiz_greens_2015,martin-ruiz_electro-2016_alt,martin-ruiz_electromagnetic_2016_alt,bonilla_electromagnetic_2023,barredo-alamilla_axion_2024}. This approach has been used to compute different observables for systems with a $\theta$ discontinuity, i.e., in the presence of $\theta$ boundaries. 

This article introduces an alternative method to the study of axion topological materials based on the use of surface variables. This method has proven to be a successful strategy for many physical problems. For example, the computation of electric forces on charges in systems with different permittivities is a necessary step during molecular simulation of ionic solutions, charged macromolecules and colloids \cite{nguyen_incorporating_2019}. This computation requires the solution of the Poisson equation in domains with non-uniform permittivity. The problem can be approached by describing the forces as composed of direct interactions between free charges and a contribution from the net polarization charge that appears the interfaces between regions with different permittivities. In this context, the surface polarization charge is the surface variable. Its value is determined by the net electric field created at the interfaces by both the free charges and the interface charges themselves. This leads to a boundary integral equation for the surface charge. This is typically solved by discretization of the surface and the use of finite element methods \cite{harrington_boundary_1989}.  In addition to the practical solution of the electrostatic problems, it has been established that the description by means of surface variables has other important properties. For example, it is possible to construct a positive-definite free energy functional in terms of surface charges \cite{jadhao_simulation_2012,jadhao_free-energy_2013}. In this paper we follow the route of surface variables to the description of static axion topological materials. We describe its electromagnetic properties using surface variables that can be identified as induced charge and current densities. Their values can be used to recover all other macroscopic bulk properties. We show that these variables satisfy a functional variational energy in the quasi-static regime, where radiation effects are absent. In this regime the electric and magnetic fields, and the charge and current distributions are static. Although currents are present, there is no charge accumulation and the charge distribution can remain constant. The charge and currents satisfy a set of coupled integral equations. These can be solved analytically for simple cases. For these cases with high symmetry we show that our results are equivalent to those using the Green function method \cite{martin-ruiz_greens_2015,martin-ruiz_electro-2016_alt,martin-ruiz_electromagnetic_2016_alt,bonilla_electromagnetic_2023,barredo-alamilla_axion_2024}. For more complex configurations they can be solved numerically by means of a boundary finite element method. The use of surface variables reduces the dimensionality of the integral equations. 

This article is organized as follows: In Sec. 2 we present the axion electrodynamics equations in the quasi-static regime and shown how they follow form a Hamiltonian formulation. Sec. 3 is devoted to the discussion of the surface variables in the case of topological insulators. in Sec. 4 we present the surface integral equations satisfied by these variables. In Sec. 5 we present analytic solutions for axial-symmetric systems, where an exact solution is given. Sec. 6 is used to discuss the general numerical methods which are implemented by discretizing the surface variables. In Sec. 7 the numerical methods are applied to find the solution of several systems of topological insulators with different geometries such as spheres, hollow spheres and toroidal surfaces. Finally Sec. 8 is devoted to discuss the final remarks and conclusions.

\section{Hamiltonian description}

To study  quasi-static systems with axion electrodynamics, we start from a Hamiltonian formulation that uses as base variables the electric $\mathbf{E}$ and magnetic fields $\mathbf{B}$. 
A suitable quasi-static Hamiltonian can be written that incorporates their coupling via the axion field $\theta$.  In the application considered, modeling topological insulators, the dielectric and paramagnetic constants can be taken to be those of the vacuum. The effective  dynamics of the axion field is not specified and it is assumed to maintain its constant value inside the topological insulator. We include a simple matter term to recover the correct equations for the fields but omit further terms that provide the forces keeping the charges and currents in place.   We use a Hamiltonian density that depends, on the fields, their potentials and on the position and momentum of charged particles  $\mathbf{x}$, $\mathbf{p}$. The particles interact with the fields only through the net charge and current densities, $\rho$ and $\mathbf{j}$. The particles produce a particle momentum field  density $\mathbf{p}$. The Hamiltonian density is    
\begin{eqnarray}
\mathcal{H} & = & \frac{1}{2}(\mathbf{E}^{2}-\mathbf{B}^{2})+\theta \mathbf{E}\cdot\mathbf{B}+\frac{1}{2m}(\mathbf{p}-\rho\mathbf{A}\big)^2\nonumber\\&&-\phi (\nabla\cdot(\mathbf{E}+\theta\mathbf{B})-\rho) \nonumber\\
 &  & +\mathbf{A\cdot}\nabla\times \big(\mathbf{B}-\theta\mathbf{E}).
\end{eqnarray} 
The third term is the kinetic contribution of a continuous density of free charges. For simplicity we only consider a single species with  $m$ its mass density and $\mathbf{p}$ its momentum density, and omit potential energy terms. We use units where the permittivity and permeability of the vacuum have a numerical value of 1. Relative permittivity and permeability of the insulator material are also set to $1$. 

In a Lagrangian formulation, the  electric and magnetic potentials $\phi$ and $\mathbf{A}$ can be introduced as Lagrange multipliers that enforce the relations between free charge $\rho$ and current $\mathbf{j}$ sources and the fields. Passing to a Hamiltonian formulation, the electric potential and magnetic field become auxiliary variables while the magnetic potential can be identified with the momentum of the electric field. Details of this construction appear in the  appendices. 

The quasi-static condition is that all variables must not have time dependence. The system contains currents composed of moving charges but the net currents maintain a constant  value.  The Hamiltonian equations for conjugate variables $Q$ and $P$ have the form
$\dot{Q}=\partial_{P}H,\,\,\dot{P}=-\partial_{Q}H$ and, therefore, in the quasi-static case, the equations for the fields amount to the extremization of the Hamiltonian with respect to the base variables and their momenta. The matter field still gives rise to position and velocity time derivatives as currents are present. There are no effective momentum equations associated to $\mathbf{B}$, and $\phi$, as these variables do not have  kinetic terms. The Hamiltonian is simply extremized with respect to these variables. 

The Hamiltonian equation for the matter positions ($\partial_t\mathbf{x}=\partial_\mathbf{p} H$) identifies the velocity distribution $\mathbf{v}$ as
\begin{equation}
    \mathbf{v}=(\mathbf{p}-\rho\mathbf{A})/m. 
\end{equation}
The current density $\mathbf{j}$ is then written as
\begin{equation}
    \mathbf{j}=\frac{\rho}{m}(\mathbf{p}-\rho\mathbf{A}). 
\end{equation}
The ratio of charge to mass densities is just the charge to mass ratio for a single species. Variation of the Hamiltonian with respect to the particle positions is computed through the variation of charge and current densities. The time derivative of the momentum then identifies the forces required to maintain the currents in place. This result is not needed for the following but is sketched in the appendices. 

The equations obtained from the minimization of the Hamiltonian with respect to the fields are:
\begin{eqnarray}
&&\mathbf{E}+\theta\mathbf{B}+\nabla\phi-\theta\nabla\times\mathbf{A}=0,\label{eq:Eeq}\\&&
-\mathbf{B}+\theta\mathbf{E}+\nabla\times\mathbf{A}+\theta\nabla\phi=0.\label{eq:Beq}
\end{eqnarray}
These can be seen to require the standard relations between potentials and fields:
\begin{eqnarray}
\mathbf{E}&=&-\nabla\phi,\label{eq:Eeq2}\\
\mathbf{B}&=&\nabla\times\mathbf{A}.\label{eq:Beq2}
\end{eqnarray}
We note that these imply that $\nabla\times\mathbf{E}=0$ and $\nabla\cdot\mathbf{B}=0$.
Equations associated to the potentials give
\begin{eqnarray}
&&
\nabla\cdot(\mathbf{E}+\theta\mathbf{B})=\rho, \label{sourcerho}
\\
&&\nabla\times(\mathbf{B}-\theta\mathbf{E})=\mathbf{j}.
\label{sourcej}
\end{eqnarray}
In terms of the potentials these equations read
\begin{eqnarray}
&&
-\nabla^{2}\phi+(\nabla\theta\cdot\nabla\times\mathbf{A})=\rho,
\\
&&\nabla\times(\nabla\times\mathbf{A})+\nabla\theta\times\mathbf{\nabla\phi}=\mathbf{j}.
\end{eqnarray}
We note that when the equations of motion are satisfied, and if fields and potentials decay fast enough at infinity, we have after integrating by parts, 
\begin{equation}
\mathbf{\int A\cdot}[\nabla\times(\mathbf{B}+\theta\mathbf{E})]\,dV=\int [\mathbf{B\cdot B}+\theta\mathbf{B}\cdot\mathbf{E}]\,dV.
\end{equation}
Using these relations can be seen that the evaluation of the energy density $\mathcal{E}$ at equilibrium is given by expressions that coincide with the standard form:
\begin{equation}
\mathcal{E}=\frac{1}{2}(\mathbf{E}^2+\mathbf{B}^2)+\frac{1}{2}mv^2. 
\end{equation}
That is, the presence of the axion term modifies the equilibrium configurations and leads to different numerical results, but does not change the form of evaluation of the net energy density in terms of the equilibrium fields. Uniform fields that do not decay leads to other non-zero contributions.  

\section{Surface variables }

The system of eqs. (\ref{sourcerho}) and (\ref{sourcej}) for the fields and sources can be restated as 
\begin{eqnarray}
&&\nabla\cdot\mathbf{E}=\rho+\omega, \label{withomega}\\
&&\nabla\times\mathbf{B}=\mathbf{j}+\mathbf{s},\label{withj}
\end{eqnarray}
where we have introduced the induced charge and current distributions $\omega$, $\mathbf{s}$ respectively:
\begin{eqnarray}
&&\omega=-\nabla\theta\cdot\mathbf{B}, \label{defomega}\\
&&\mathbf{s}=\nabla\theta\times\mathbf{E}.\label{defj}
\end{eqnarray}
For piece-wise continuous axion fields, the charge and current densities are localized at the interfaces between regions with constant axion values. We will therefore refer to these induced sources as surface variables.   

In terms of potentials these equations are equivalent to 
\begin{eqnarray}
&&\nabla^{2}\phi=-(\rho+\omega), \label{poissonrho}\\
&&\nabla^{2}\mathbf{A}=-(\mathbf{j}+\mathbf{s}),\label{poissonj}
\end{eqnarray}
where the last equation is obtained after adding the condition $\nabla\cdot\mathbf{A}=0$, which does not modify results for the magnetic field. 
We assume that all matter and all interfaces are present only in a finite region of space. Then, the solution for the potentials can be written
in terms of the sources as 
\begin{eqnarray}
&&\phi=\int G_{xy}(\rho+\omega)_{y}dV_{y} \label{phisol},\\
&&\mathbf{A}=\int G_{xy}(\mathbf{j}+\mathbf{s})_{y}dV_{y}.\label{potAsol}
\end{eqnarray}
 In this expression we use subindices $x,y$ to indicate the position variables. The geometric Green's function for free space is
\begin{equation}
    G_{xy} \equiv G(\mathbf{x},\mathbf{y})=\frac{1}{4\pi}\frac{1}{|\mathbf{x}-\mathbf{y}|}. \label{Greens}
\end{equation}
The fields are:
\begin{eqnarray}
&&\mathbf{E}=-\nabla\int G_{xy}(\rho+\omega)_{y}dV_{y}, \label{Esol}\\
&&\mathbf{B}=\nabla\times\int G_{xy}(\mathbf{j}+\mathbf{s})_{y}dV_{y}.\label{Bsol}
\end{eqnarray}
To these expressions one can add constant fields $\mathbf{E}_{0}$, and $\mathbf{B}_{0}$.  They can be excluded here since when the induced sources are localized in a finite region, the effect of constant fields can be mimicked by using large but finite distant charge and current sources (a very large capacitor and coil). We restore explicit mention of the constant external fields in the numerical examples.  

Using the expressions for the fields in terms of free and induced sources, we obtain 
\begin{eqnarray}
\omega & = & -\nabla\theta\cdot \left(\nabla\times\int G_{xy}(\mathbf{j}+\mathbf{s})_{y}dV_{y}\right),\label{omegaEq}\\
\mathbf{s} & = & \nabla\theta\times \left(-\nabla\int G_{xy}(\rho+\omega)_{y}dV_{y}\right).\label{sEq}
\end{eqnarray}
These are the integral equations satisfied by the surface variables. This is the key result of this approach. In the sections below we show how these equations can be effectively solved numerically and the results can be used to reconstruct the values of the fields at all points. 

We can now construct a new Hamiltonian that uses the induced  charge and current, $\omega$ and $\mathbf{s}$ as base variables.  A suitable quasi-static Hamiltonian is 
\begin{eqnarray}
\int \mathcal{H}(\mathbf{s},\omega, \mathbf{j},\mathbf{p}) dV &&= \int\int\int H_{xyz}^{(3)} dV_xdV_ydV_z\nonumber \\ &+&\int\int H^{(2)}_{xy} dV_xdV_y+\int H^{(1)}dV_x  
\end{eqnarray}
with the integrands
\begin{eqnarray}
H_{xyz}^{(3)}&=&\theta_x \nabla_x G_{xz}(\rho+\omega)_z  \nonumber\\
&& \times \big[ \nabla_xG_{xy}\times(\mathbf{j}+\mathbf{s})_{y}\big], \\
H_{xy}^{(2)}&=&\frac{1}{2}\rho_{x}G_{xy}\rho_{y}-\frac{1}{2}\omega_{x}G_{xy}\omega_{y} \nonumber \\
 &  & +\frac{1}{2}\mathbf{j}_{y}G_{xy}\cdot \mathbf{j}_{y}+\frac{1}{2}\mathbf{s}_{y}G_{xy}\cdot \mathbf{s}_{y},\\
H_{x}^{(1)} &= &-\frac{m}{\rho^2}\mathbf{j}^2+\frac{1}{\rho}\mathbf{j}\cdot\mathbf{p}.
 \label{hamiltoniansurface}
\end{eqnarray}
This expression can be obtained by repeated application of the process of expressing the original field variables in terms of the induced densities by means of Lagrange multipliers, solving for the multipliers and eliminating them. This process results in many quadratic terms to appear with negative signs. This is a common feature when the type of variables employed is changed \cite{solis2013generating}. As noted below, however, the result coincides with the standard energy evaluation when the equation of motion are satisfied. We also note that the charge density in the denominators simply changes the currents into velocities. 

The equations for the induced charge and current follows from minimization of the functional with respect to their values. Variation with respect to the induced charge density gives
\begin{equation}
\int G_{xy}\bigg[-\omega_x-\nabla_x\theta\cdot\nabla_x\times \int G_{xy}(\mathbf{j}+\mathbf{s})_ydV_{y}\bigg]dV_x=0.
\end{equation}
The equation is satisfied when the factor in square brackets in the integrand is zero, which is eq. (\ref{omegaEq}). Similarly, variation with respect to the induced current leads to 
\begin{equation}
\int G_{xy}\bigg[\mathbf{s}_x+\nabla_x\theta\times\nabla_x\int G_{xy}(\omega +\rho)_ydV_{y}\bigg]dV_x=0,
\end{equation}
which holds when eq. (\ref{sEq}) is satisfied. 
In this formulation, the particle current has also been introduced as an auxiliary variable and variation with respect to it leads to 
\begin{equation}
(m/\rho)\mathbf{j}-\mathbf{p}+\rho\int G_{xy}(\mathbf{j}+\mathbf{s})dV_y=0.
\end{equation}
Therefore, the equations obtained from the surface variables Hamiltonian are equivalent to the ones obtained from the original expression in terms of fields. 

The quadratic expressions that appear in the original Hamiltonian can be written in terms of net charges and currents. 
After rearrangement of the terms in the surface variable Hamiltonian, it can be shown that the evaluation of the energy at equilibrium is given by 
\begin{eqnarray}
    \mathcal{E} &=&\frac{1}{2}\int\int(\rho+\omega)_{x}G_{xy}(\rho+\omega)_{y}dV_{x}dV_{y} \nonumber \\ && + 
\frac{1}{2}\int\int(\mathbf{s}+\mathbf{j})_{y}G_{xy}\cdot(\mathbf{s}+\mathbf{j})_{y}dV_{x}dV_{y} \nonumber \\ &&
+\int \frac{m}{\rho^2}\mathbf{j}^2 dV_x.\label{EnergySources}
\end{eqnarray}

\section{Surface equations}

We now  assume that the axion field is piece-wise constant. The gradient $\nabla\theta$ becomes a  delta function located at the boundary between
regions and we can write it as 
\begin{equation}
\nabla\theta=\delta(x_{n})\Delta\theta\mathbf{n}, 
\end{equation}
where $x_{n}$ is a coordinate that extends perpendicularly form the
surface and $\mathbf{n}$ is the normal to the surface. The change in the axion field value across the interface is $\Delta \theta$. We now define surface densities $\omega_s$ and $\mathbf{s}_s$ that exclude the $\delta$ function factors. That is, $\omega=\omega_s\delta(x_n)$ and $\mathbf{s}=\mathbf{s}_s\delta(x_n)$.  These satisfy, at points in a $\theta$-boundary, 
\begin{eqnarray}
\omega_s & = & -\Delta\theta\mathbf{n}\cdot\mathbf{B},\\
\mathbf{s}_s & = & -\mathbf{E}\times\mathbf{n}\Delta\theta.
\end{eqnarray}
Similarly, we can then introduce the direct induced
surface terms $\omega_D$ and $\mathbf{s}_D$:
\begin{eqnarray}
\omega_{D}&=&-\Delta\theta\mathbf{n}\!\cdot \bigg[\nabla\times\!\int G_{xy}\mathbf{j}dV_{y}+\mathbf{B}_0\bigg],\\
\mathbf{s}_{D}&=&\Delta\theta\mathbf{n}\!\times \bigg[-\nabla\!\int G_{xy}\mathbf{\rho}dV_{y}+\mathbf{E}_0\bigg],
\end{eqnarray}
where we have included a possible contribution from constant fields. These direct contributions exclude the delta function associated to the surface where they are evaluated. 
The integral equations now take the form 
\begin{eqnarray}
\omega_{s} & + & \Delta\theta\mathbf{n}\cdot\left[\nabla_x\times\int G_{xy}\mathbf{s}_ydS_{y}\right]=\omega_{D},\\
\mathbf{s}_{s} & + & \Delta\theta\mathbf{n}\times\left[\nabla_x\int G_{xy}\omega_{s\,y} dS_{y}\right]=\mathbf{s}_{D}.
\end{eqnarray}
We note that these are surface integral equations and that they are linear in the surface variables. We can schematically write
\begin{equation}
\mathbf{w+X}\mathbf{w}=\mathbf{w}_{D},
\label{opEqn}
\end{equation}
where $\mathbf{w}$ stands for the values of the surface charge and current variables while  $\mathbf{X}$ is a linear integral operator that acts on the $\mathbf{w}$. Namely,
\begin{equation}
\mathbf{w}=\left(\begin{array}{c}
\omega_{s}\\
\mathbf{s}_{s}
\end{array}\right), \label{eq:defw}
\end{equation}
and a similar structure $\mathbf{w}_D$ for the terms associated to the direct effect of the external field. The operator is 
\begin{equation}\label{eq:schematicInt}
\mathbf{X}=\int\left(\begin{array}{cc}
0 & \Delta\theta\mathbf{n}\cdot\nabla\times G\\
\Delta\theta\mathbf{n}\times\nabla G & 0
\end{array}\right)[\cdot]dS_y.
\end{equation}
The symbol $[\cdot]$ stands for the space where the vector $\mathbf{w}$ is to be inserted. We omitted the position subindices in this expression. 
This abstract form of the equations can now be easily discretized using finite boundary elements and thus reduced to a linear algebraic equation. 

After solving for the induced sources, the fields at all points in space can be obtained using the geometric Green's function. 

\section{Analytic Solutions }\label{analytic}

For simple geometries is possible to solve analytically for the induced sources and fields. This section works out the case of spherical geometries. 

\subsection{Axial symmetric systems and the single charge}

Consider the case of a spherical system in the presence of an axial symmetric external field, such as the one created by a  single charged particle. 
A sphere of radius $R$ has an axion field with
value $\theta$ while the exterior has zero field. Assume all other
properties, such as permittivity and conductivity are equal. 

In this case the axial symmetry allows us to write the surface charge and
current densities as a series in Legendre polynomials and associated
Legendre polynomials. The surface charge density is written as 
\begin{equation}
\omega_s=\sum_{n}\omega_{n}P_{n}(\cos \theta_s),
\end{equation}
where the polar angle is denoted $\theta_s$ to avoid confusion with the axion field. The axial component of the current is 
\begin{eqnarray}
\mathbf{s}_{s,\phi}&=&\sum_n s_{n}P_{n}^{1}(\cos \theta_s)\mathbf{e}_{\phi}.  \label{sPoly}
\end{eqnarray}
We use the spherical unit vectors $\mathbf{e}_r$, $\mathbf{e}_\theta$, $\mathbf{e}_\phi$ and standard spherical coordinates.  In these expressions $P_n$ are the Legendre polynomials  and $P^m_n$ are the associated Legendre polynomials. The sum for the charge is over all non-negative integers while for the current is over all positive integers.  For individual contributions to the expansions of the surface charge and currents, the application of the integral operator leads to the following evaluations:
\begin{equation}
\int \Delta\theta\mathbf{n}\times\nabla G \omega_n P_{n}dS=\frac{\Delta\theta\omega_{n}}{2n+1}P_{n}^{1}\mathbf{e}_{\phi},
\end{equation}
\begin{equation}
\int\Delta\theta\mathbf{n}\cdot\nabla\times G{s}_{n}P_n^{1}dS=-\frac{n(n+1)\Delta\theta{s}_{n}}{(2n+1)}P_{n}.
\end{equation}
These expressions are derived from the properties of the Legendre polynomials and, in particular, $\partial_{\theta}{P}_{n}(\cos \theta_s)=P_n^{1}(\cos \theta_s)$ and $\partial_\theta[\sin \theta_s P_{n}^{1}(\cos \theta_s)]=-(\sin\theta_s)n(n+1)P_n(\cos \theta_s)$. We obtain a set of equations for the amplitudes:
\begin{eqnarray}
\omega_{n}- \Delta\theta \frac{n (n+1)}{2n+1}s_{n}=\omega_{D,n},\\
s_{n}+\Delta\theta \frac{1}{2n+1}\omega_{n}=s_{D,n}.
\end{eqnarray}
Solutions of the system are given by 
\begin{eqnarray}
\omega_{n}=\frac{1}{D_n} \left[\omega_{D,n}+\Delta\theta\frac{n(n+1)}{(2n+1)}s_{D,n}\right],\label{solwn}\\
s_{n}=\frac{1}{D_n} \left[-\frac{\Delta\theta}{(2n+1)}\omega_{D,n}+s_{D,n}\right],\label{solsn}
\end{eqnarray}
with
\begin{equation}
D_n=1+n(n+1)\Delta\theta^2/(1+2n)^2.\label{solD}
\end{equation}

If a single charge is at location $(x,y,z)=(0,0,a)$ outside the sphere, it creates a potential 
\begin{equation}
\phi=\sum_n\frac{R^{n}}{4\pi a^{n+1}}P_{n}(\cos\theta_s)
\end{equation}
at the surface of the sphere. The field due to the charge is 
$
\mathbf{E}=\mathbf{E}_{r}+\mathbf{E}_{\theta}$
with components 
\begin{eqnarray}
\mathbf{E}_{r}=-\sum_n\frac{R^{n-1}}{4\pi a^{n+1}}nP_{n}(\cos\theta)\mathbf{{e}}_r,\\
\mathbf{E}_{\theta}=
-\sum_n\frac{R^{n-1}}{4\pi a^{n+1}}P_{n}^{1}(\cos\theta){\mathbf{e}}_{\theta}.
\end{eqnarray}
The surface current generated by the external charge is written as
\begin{equation}
\mathbf{s}_{D}=\sum_{n}s_{D,n}P_{n}^{1}(\cos \theta)\mathbf{e}_{\phi},
\label{sourceone}
\end{equation}
with 
\begin{equation}
s_{D,n}=-\Delta\theta\frac{R^{n-1}}{4 \pi a^{n+1}}.
\label{sourcetwo}
\end{equation}
The pure external electric field does not produce a direct surface charge
so that $\omega_{D}=0$. 

The net induced surface charges can now be read from the solutions eqs. (\ref{solwn}), (\ref{solsn}) and (\ref{solD}). 

The electric and magnetic fields are easily reconstructed from these values. The electric and magnetic potentials due to the induced sources $\phi_s$ and $\mathbf{A}_s$ are
\begin{equation}
\phi_{s}=\sum_n \frac{R^2}{(2n+1)}\omega_n \frac{r_<^n}{r_>^{n+1}}P_n(\cos \theta_s) 
\label{scalarf}
\end{equation}
and 
\begin{equation}
\mathbf{A}_{s}=\sum_n \frac{R^2}{(2n+1)}s_n \frac{r_<^n}{r_>^{n+1}}P^1_n(\cos \theta_s) \mathbf{e}_\phi.
\label{vectorf}
\end{equation}
In these expressions, $r_<$ is the smaller of the radial observation position $r$ and the sphere radius $R$, while $r_>$ is the largest of these two distances. The results coincide with those in ref. \cite{martin-ruiz_electro-2016_alt}. 
 
\subsection{External field}

A second example with axial symmetry is provided by the application of  external constant fields $\mathbf{E}_0$ and $\mathbf{B}_0$ aligned with the $z$-axis. The axion topological matter is again spherical, of radius $R$. The direct effect contribution has a dipolar shape and thus only $\omega_{D,1}$ and $s_{D,1}$ are not zero. 
Their values are 
\begin{eqnarray}
\omega_{D,1}=-B_0\Delta\theta, \\
s_{D,1}=E_0\Delta\theta. 
\end{eqnarray}

The surface densities are read again from the eqs. (\ref{solwn}), (\ref{solsn}) and (\ref{solD}). Their only non-zero term has index $n=1$ and the solution can be explicitly written as:
\begin{eqnarray}
\omega=D^{-1}\bigg(\frac{2}{3}\Delta\theta^{2}E_{0}-\Delta\theta B_{0}\bigg)\cos\theta_s,\\
\mathbf{s}=-D^{-1}\Bigg(\Delta\theta E_{0}+\frac{1}{3}\Delta\theta^{2}B_{0}\bigg)\sin\theta_s\mathbf{e}_{\phi},
\end{eqnarray}
with $D=1+(2/9)\Delta\theta^2$. From these values, the potentials and total fields can be constructed. 



\section{Numerical Methods}

The schematic integral equation (\ref{eq:schematicInt}) can be solved numerically by discretizing the surface variables. We retain the same notation for the discrete version of the system. We assume that there are $S$ surfaces where there is a jump in the axion field value. The surfaces are triangulated into a total of  $N_T$ triangles. The triangle centers are $\mathbf{x}_{t}$, defined by the average of their vertices. The normal to each triangle is a unit vector $\mathbf{n}_{t}$. A surface charge density $\omega_t$ and a current density vector $\mathbf{s}_{t}$ are assigned to each triangle. The discretized version of the vector $\mathbf{w}$ is a column vector of length $4N_t$, with the structure:
\begin{equation}
    \mathbf{w}=(\omega_1,\ldots,\omega_{Nt},s_{1,1},s_{1,2},\ldots,s_{N_{t,3}})^T,
\end{equation}
where $s_{t,i}$ are the cartesian components ($i=1,2,3$) of the current at triangle $t$. 
The superscript $T$ denotes transposition so as to obtain a column vector. We also write, in block notation,
\begin{equation}
    \mathbf{w}=(\omega_1,\ldots,\omega_{Nt},\mathbf{s}_1,\ldots,\mathbf{s}_{N_t})^T.
\end{equation}
The discretized vector $\mathbf{w}_D$ is constructed in identical fashion. 

The application of the operator $\mathbf{X}$ reduces to a matrix multiplication. Its discretization becomes a square matrix of size $4N_t$. It can be expressed in block structure as
\begin{equation}
X=\left(\begin{array}{cc}
0 & X_B\\
X_C & 0
\end{array}\right).
\end{equation}
The matrix $X_B$ can be expressed as composed of $N_t\times N_t$ blocks. In this form, the block with indices $(t,t')$, $t\neq t'$, is a horizontal vector $B^{(t,t')}$ of size $1\times 3$ with entries $B^{(t,t')}_j$, for $j=1,2,3$ given by
\begin{equation}
B^{(t,t')}_i=\epsilon_{ijk}n^{(t)}_j\partial^{(t)}_kG^{(t,t')} S^{(t')}.
\end{equation}
That is, this vector is an approximation to the integrand $dS_{t'}\mathbf{n}\cdot \nabla^{(t)}\times G(\mathbf{x}_t,\mathbf{x}_{t'})[\cdot]$. In this expression, $\epsilon_{ijl}$ is the alternating symbol and $S^{(t)}$ is the area of the $t$-th triangle. This  results from assuming that the current in the $t'$ triangle is constant and that the value of the Green's function can be approximated by its evaluation with a source at the center of the triangle.  The partial derivative is taken with respect to the location of the center of the $t$ triangle. 
Similarly, the matrix $X_C$ as composed of $N_t\times N_t$ blocks. These blocks are column  vectors $C^{(t,t')}$ of size $3\times 1$ that are the transpose of the  $B^{(t,t')}$ blocks. 
The terms with equal indices $t=t'$ are set to zero, as the electric and magnetic fields associated with flat finite elements with uniform surface densities can be shown to have zero normal components at surface points.

The  choices for approximation of the surface distributions and the operators described above require some refinement for effective applications. Empirically we find two necessary improvements. The scheme above evaluates the Green's function at the center of the triangles. For triangles very close to each other is better to replace the one point evaluation by the actual integration over the surface. We employ this refinement for pairs of triangles at distances less than a few triangle size. The integration of the Green function can be completed analytically and the resulting integral used. The second refinement is a smoothing of the charge and current values obtained by averaging over a few neighboring triangles. With these modifications, the method recovers known analytical solutions and produces sensible results for generic cases. 

To solve the equation we use an iterative procedure. We evaluate the fields created by the free sources in free space and determine their contribution to the surface charge and current distributions; the vector $\mathbf{w}_D$ is therefore given. We use this as our first approximation: $\mathbf{w}^{[0]}=\mathbf{w}_D$. We compute iterations of the vector $\mathbf{w}'$:
\begin{equation}
\mathbf{w'}^{[i+1]}=\mathbf{w}_D-\mathbf{X}\mathbf{w}^{[i]}.
\end{equation}
We find that directly using this vector as the next step in the approximation has numerical instabilities. These instabilities are avoided by a smoothing step, schematically indicated as
\begin{equation}
\mathbf{w}^{[i+1]}=S\mathbf{w'}^{[i+1]}.
\end{equation}
The smoothing step is a linear operation where the values of the charge and current at triangle $t$ are set to the average over the value determined by $\mathbf{w}'$ at that triangle along with its three side-neighbors. This process amounts to a smoother approximation to the interaction operator $\mathbf{X}$. 

The iterative process stops when the new average magnitude of the density differs from the previous value by a small fraction ($10^{-3}$ in the numerical examples below). The most interesting cases have a relatively small contrast values $\Delta \theta$ and we observe very quick convergence.

Once the surface charges and currents are determined, the fields at all points of space can be reconstructed by using the total charges and net currents in the triangular  elements as point sources. Their contributions to the net field are given by their Coulomb and Biot-Savart fields. However, for locations closer to the surface, the contribution to the total field of nearby segments is better approximated by uniform fields. Very near a surface element with surface charge density $\omega$ the electric field is $\omega/2$ and perpendicular to the triangle surface. The magnetic field is parallel to the surface, perpendicular to the current, and of magnitude $|\mathbf{s}|$. 
\section{Numerical examples}

We apply the computational method to examples with different geometries. To establish the validity of the method we consider a spherical case with a prescribed external field. As the solution to this case can be obtained analytically, it provides a concrete measure of the method accuracy. As an example of a system with non-trivial topology we consider a torus in the presence of external uniform fields. Finally, we show that the method can be applied to multiple disconnected domains and consider the case of a hollow sphere. 

We use dimensionless variables. All distances are measured with respect to a characteristic length $L$, electric fields are measured as multiples of $e/(\varepsilon_0L^2)$, while magnetic fields as  multiples of $e/(c\varepsilon_0L^2)$, with $c$ the speed of light. The axion field can be measured in multiples of the speed of light. For the concrete examples below we take $\theta_0=0.0073=\approx 1/137$ (the fine-structure constant). In the examples below, the material is considered a topological insulator with unit relative permittivity and permittivity. All induced currents and charges appear only at the surface between the axion topological material and its environment.

\subsection{Spherical geometries}
To validate the method we consider two families of systems based on spherical geometries.  The sphere contains a material described by a constant axion field, with value $\theta_0=0.0073$. The field vanishes outside the sphere and thus has a jump of magnitude $\Delta \theta=-\theta_0$ at the sphere surface. In the presence of external fields, created outside the sphere, the surface acquires a surface charge distribution and a  surface current distribution. We consider the cases where the external fields applied are constant, and when the field is generated by a point source. These correspond to the cases studied by analytic methods in sec. \ref{analytic}. In both cases the sphere is discretized into 1280 triangles obtained from a triangular refinement of an icosahedron. The radius of the sphere sets the scale of the system and thus is taken to be $R=1$. 

\begin{figure}
\includegraphics[width=3in]{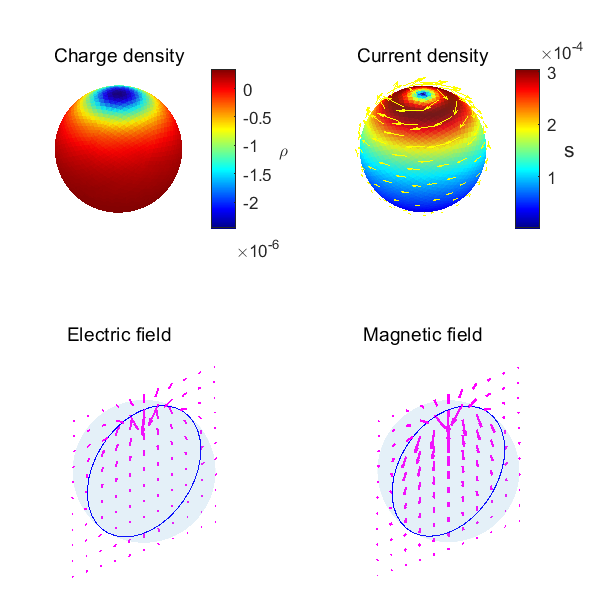}
\caption{
Response of a  topological insulator to the field of a point charge. Using dimensionless quantities, the sphere radius is $R=1$. An unit charge is located at $z=2$. Top panels show the induced surface charge and current densities. Bottom panels show electric and magnetic field maps evaluated at a vertical plane through the sphere center. These fields are created by the induced charge and current and do not include the external field. }\label{fig:sphere2} 
\end{figure}

In the case of a constant external electric or magnetic field, aligned with the $z-$axis, the resulting charge distribution is proportional to $\cos\theta_s$, the first spherical harmonic. The surface current  circulates around the $z-$axis. Numerical solutions obtained agree with the analytical result to less than $1.0\%$ error. For a second spherical example, the material is in the presence of charge points of unit magnitude. The charge is placed at $(x,y,z)=(0,0,2)$. The resulting current and charge distributions are axial-symmetric. The solutions agree with the analytical result to less than $1.0\%$ error. The results for this second case are shown in fig. \ref{fig:sphere2}.  In this and subsequent figures we show, in the top left,  the surface charge density as a color map and in the top right the magnitude of the surface current density as a color map. In addition, the direction of the surface current is shown by an arrow map field. The current is parallel to the surface. The arrows are displaced from the surface for visibility. Maps of the electric and magnetic fields created by these distributions are shown in the bottom panels. Note that the fields shown do not include the external contribution. The fields shown are evaluations at the vertical plane $x=0$.  A translucent version of the surface is included in these figures and its intersection with the plane is shown with solid lines. The scale of the fields is chosen for visibility.

In the case of a point charge,  fig. \ref{fig:sphere2} shows that the main effect of the external charge is the creation of a surface current density. The surface charge is induced by the resulting magnetic fields but its magnitude is much smaller. The induced currents are larger near the north pole as the region is closer to the point source. Due to the surface charges and currents the induced fields inside an outside the material are not continuous. 

\subsection{Toroidal surfaces}
Our numerical method can be used for non-trivial geometries. We consider next the case of a topological insulator with toroidal shape. In all the examples presented, the torus is obtained from revolution, around a circular path of radius $R_1=1$, of a solid circle of radius $R_2=0.5$. Its axis of symmetry is the $y$-axis. The discretization of the torus is obtained by deforming a cylinder constructed from a triangular lattice. Examples below use 3321 triangles in the torus discretization. 

\begin{figure}
\includegraphics[width=3in]{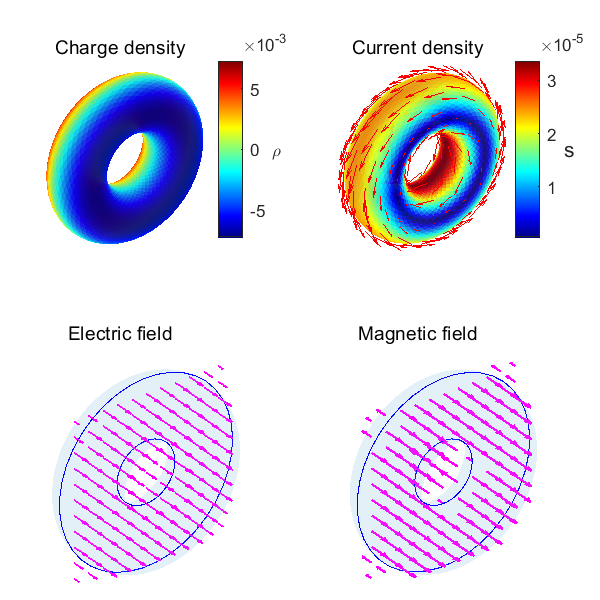}
\caption{
Response of a toroidal shape topological insulator to a uniform external magnetic field along the $y$-axis. The $y$-axis is aligned with the axis of symmetry of the torus and points towards the back of the figure.  }\label{fig:torusBY} 
\end{figure}

As a first example, we apply a uniform magnetic field along the $y$-axis. Results are shown in fig. \ref{fig:torusBY}. As the external field is magnetic, the response is dominated by the surface charge directly induced by the field at the surface of the torus. Secondary effects modify this surface charge density and produce an induced current density. At the mid-plane section, inside the torus, the  magnetic field created by the induced current points in the direction opposite to the external field. 

\begin{figure}
\includegraphics[width=3in]{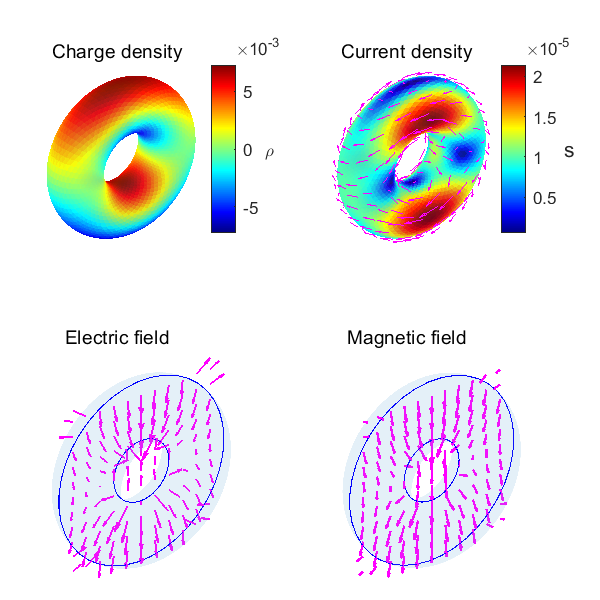}
\caption{
Response of a toroidal shape topological insulator to a uniform external electric field along the $z$-axis. The $z$-axis points up in the space of the figure. } \label{fig:torusEZ} 
\end{figure}

Next, we apply a uniform electric field along the $z$-axis. Results are shown in fig. \ref{fig:torusEZ}. In this case the response is dominated by the surface current directly induced by the external field. Secondary effects modify this current. The magnetic field created by the induced current results on an induced surface charge density. At the mid-plane section, all secondary fields are parallel to the plane. 

\begin{figure}
\includegraphics[width=3in]{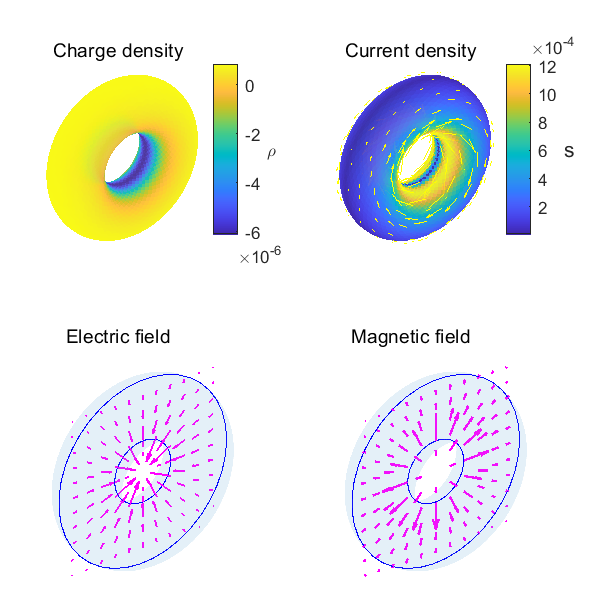}
\caption{
Response of a toroidal shape topological insulator to the presence of a unit charge at its center. The induced current circulates in opposite directions in the front and back of the torus.  } \label{fig:torusQsource} 
\end{figure}

Placing a unit charge at the center of the torus creates electric fields that, at most positions on the surface, have non-zero cross product with the surface normal and thus produce a surface current density. Interaction with the magnetic field created by this current creates a surface charge distribution.   Results for this case are shown in fig. \ref{fig:torusQsource}. We note that the net charge of the surface is zero. However, the local density is negative and high in magnitude near the center and positive but small over a larger area away from the center. 

The method can be applied to cases with any combination of uniform electric and magnetic fields. Electric and magnetic point sources can also be placed anywhere in the system though locations too close to the surface can result in larger numerical errors.  

\subsection{Multiple surfaces}

Our numerical method can address axion topological material shapes with multiple disconnected components or multiple disconnected surfaces. We present two cases of a spherical system with a spherical hollow region.

First we consider a finite version of the geometry associated to the creation of dyons (Witten's effect) \cite{witten_dyons_1979_alt}. A magnetic monopole is placed in the center of the hollow region. The hollow region is centered inside the spherical volume; the axion field value is zero there. In this case there are two discretized spherical surfaces. We use $R_1=1$ for the external radius and $R_2=0.4$ for the internal one. We use for each the same triangulation as in the single sphere case. The resulting net fields are known: in addition to the radial magnetic field, a radial electric field appears in the interior of the insulator. The fields inside the insulator thus mimic the presence of a dyon, a particle with both magnetic and electric charges in a non-axion medium. 

\begin{figure}
\includegraphics[width=3in]{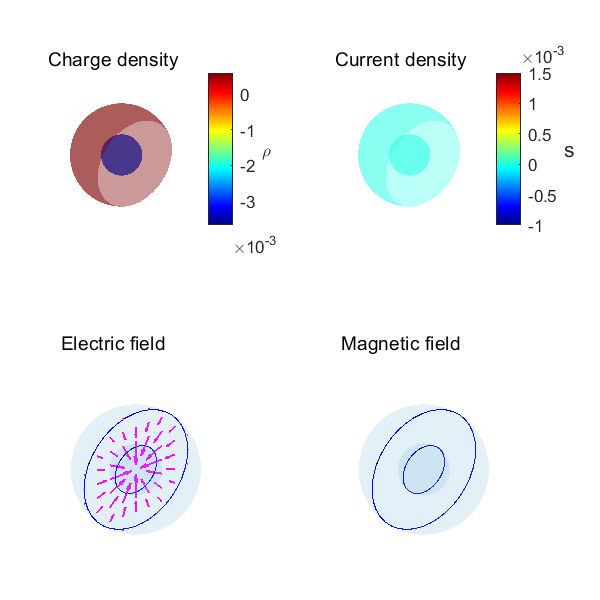}
\caption{
Response of a hollow spherical insulator to the presence of a unit monopole charge at its center. Only induced surface charges appear at the boundaries of the spheres. There are no induced currents. } \label{fig:fig5dyon} 
\end{figure}

The numerical solution to the system recovers the expected behavior. As shown in fig. \ref{fig:fig5dyon}, a uniform surface charge density is generated at the interior surface. This charge produces a radial electric field. At the exterior surface the radial magnetic field produces a surface charge. The  sign is opposite as the axion field gradient points in the opposite direction with respect to the inner surface. The net charge in the system is zero and there is no electric field outside the external surface. The magnetic field is due only to the monopole. Due to the symmetry of the system, there is no induced surface current. 

\begin{figure}
\includegraphics[width=3in]{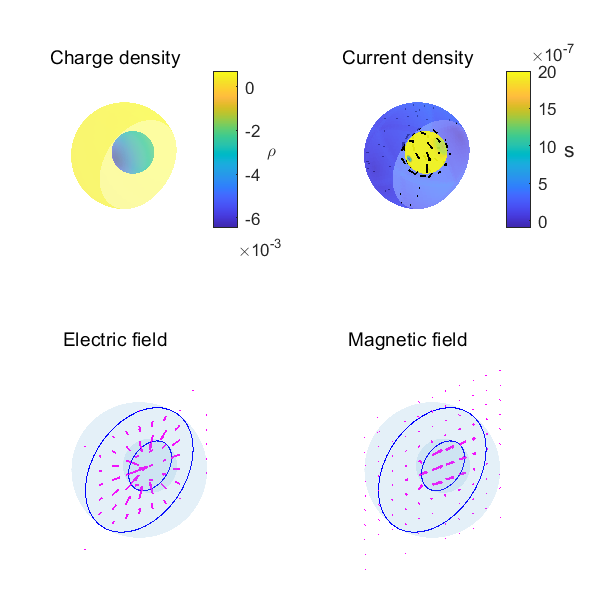}
\caption{
Response of a hollow spherical insulator to the presence of a unit monopole charge. The hollow region is off center and the charge is not located at the center of this region.  Induced surface charges and currents appear at the boundaries of the spheres.  } \label{fig:fig6dyonoffcenter} 
\end{figure}

The previous example is sensitive to geometric details. To show this we modify the geometry by setting the center of the hollow region at $(x,y,z)=(0.2,0,0)$ and the monopole at $(x,y,z)=(0.1,0,0)$. The magnetic field is not perpendicular to the sphere internal and external surfaces and generates a non-uniform charge distribution. Secondary fields do appear in the solution. As shown in fig. \ref{fig:fig6dyonoffcenter}, surface currents appear in both surfaces. The charge and current densities are larger in the interior surface but also appear in the external surface. The net charge is again zero but electric fields do appear in this case outside the surface. 

\label{S-FinalRemarks}

\section{Conclusions}
In this article, we presented an analysis of the axion electrodynamics in terms of surface variables eqs. (\ref{omegaEq}) and (\ref{sEq}). These variables were previously used in various problems of soft condensed matter. 

We have used a Hamiltonian formulation as starting point. This approach is useful as it connects more directly with energy considerations and determinations of forces when parameters of the problem are modified. Most previous discussions of systems with axion electrodynamics have been carried using action formulations or directly through the modified Maxwell equations. 
Moreover, for simplicity we restrict ourselves the analysis to time-independent phenomena in the quasi-static regime. 
We have presented the modified Maxwell equations in terms of surface variables and the integral eqs. (\ref{opEqn}), (\ref{eq:defw}) and (\ref{eq:schematicInt}) that these variables satisfy. 

For  topological insulators with axial symmetry and simple enough geometries, the surface equations, described by the integral equations, do admit analytic solutions. In Sec. 5 we obtain explicit exact solutions given by eqs. (\ref{scalarf}) and (\ref{vectorf}) for the fields and by eqs. (\ref{sourceone}) and (\ref{sourcetwo}) for the surface variables. The general statement of the numerical method is later introduced and used to find numerical solutions for topological insulators with various geometries and topologies. Examples with constant external fields and those generated by point charges outside the axion topological material are provided. In particular the case of a point electric charge outside the topological insulator with spherical symmetry is discussed as a procedure to calibrate the numerical method. After that, the case of toroidal topological defects with electric and/or magnetic charges outside the torus but placed in the center of the torus is analyzed. It is clear that a combination of these situations with electric and magnetic fields  or with external electric or magnetic point charges can be easily implemented by linear superposition. Finally, the situation of multiple surfaces can be also studied. Here the case of a magnetic monopole at the center of a  hollow sphere is analyzed.  Moreover it is also observed that if the magnetic monopole has a small displacement from the center of the inner sphere, in this situation the symmetry condition is broken and the numerical method provides the asymmetric distribution of the fields for this system. 

The incorporation of the the methodology for more realistic materials with permittivities and permeabilities as studied in \cite{Martin-Ruiz:2016joy} can be carried over in a straightforward way since the method of surface variables was originally used precisely in this context \cite{jadhao_simulation_2012,jadhao_free-energy_2013}. 

It would be interesting to study by the method of surface variables the magnetoelectric response under electromagnetic external fields of a Weyl semimetal
\cite{Martin-Ruiz:2018ets}. It seem very reasonable to implement the mentioned method in this situation. The use of surface variables have been shown to be applicable to low-frequency cases where no radiation is present \cite{solis_conduction_2024}.   Moreover, it would be also interesting to extend the study to the time-dependent situation in the context of the surface variables analysis described in the present work. For instance the study of radiating systems \cite{Ibarra-Meneses:2026dul} and antennas  \cite{Ibarra-Meneses:2024qrw}
and other time-dependent phenomena using this technique will be worked out in future research. 

\section{Disclosures}

\noindent \textbf{Author contribution statement.} F. J. Solis and H. Garc\'{\i}a-Compe\'an contributed to the derivation 
of the integral equations and variational principles. F. J. Solis developed the numerical method presented.

\noindent\textbf{Competing Interests and Funding.} The authors have no relevant financial or non-financial interests to disclose. No funding was received for conducting this study.

\noindent\textbf{Data availability}.  All data employed are included in the 
article.


%


\appendix

\section{Axion Lagrangian}

The techniques developed in this article focus on the solution of the steady state problem for axion fields. It is useful, however, to connect this work to a broader context. In the appendices to this article, we derive the static Hamiltonian from a non-static Lagrangian. First, we obtain equations for the electromagnetic fields from a non-relativistic Lagrangian and construct its associated Hamiltonian formulation. The use of electric and magnetic fields as base variables in the non-axion case is considered in chapter 9 of Schwinger's text \cite{schwinger_classical_2019mod}. The present description adds an axion term and further emphasizes the role of the potentials as Lagrange multipliers. 

The Lagrangian density for the electromagnetic field with a theta-term is taken as 
\begin{eqnarray}
\mathcal{L} & = & \frac{1}{2}(\mathbf{B}^{2}-\mathbf{E}^{2})+\theta\mathbf{E}\cdot\mathbf{B}+\phi[\nabla\cdot(\mathbf{E}-\theta \mathbf{B})-\rho] \nonumber\\
 &  & +\mathbf{A}\cdot\big[\mathbf{j+}\partial_{t}(\mathbf{E}-\theta\mathbf{B})-\nabla\times(\mathbf{B}+\theta\mathbf{E})\big].
\end{eqnarray}
This particular form has the following features.  It considers the fields as fundamental variables. The axion field is not dynamic and is assumed to be known. The potentials $\phi$ and $\mathbf{A}$ are introduced as Lagrange
multipliers. 

For free charges in the system we can
consider the Lagrangian matter density
\begin{equation}
\mathcal{L}=\frac{1}{2}m\mathbf{v}^{2}+\mathcal{L}_{C},
\end{equation}
where the second term $\mathcal{L}_{C}$ is a constraint to maintain the currents
in specific paths. The charged current is   $\mathbf{j}=\rho \mathbf{v}$.

Variations of the Lagrangian density with respect to the fields give 
\begin{eqnarray}
-\mathbf{E}+\theta \ \mathbf{B}-\nabla\phi-\dot{\mathbf{A}}-\theta\nabla\times\mathbf{A}=0, \\
\mathbf{B}+\theta \ \mathbf{E}+\theta\nabla\phi+\theta \ {\mathbf{A}}-\nabla\times\mathbf{A}=0.
\end{eqnarray}
Variation with respect to the potentials simply retrieves the constraints.  These imply the usual relation between potentials and fields:
\begin{eqnarray}
\mathbf{B}=\nabla\times\mathbf{A},\\
\mathbf{E}=-\nabla\phi-\dot{\mathbf{A}}.
\end{eqnarray}
Then, we see that we still have
\begin{eqnarray}
\nabla\times\mathbf{E}=-\partial_{t}\mathbf{B},\\
\nabla\cdot\mathbf{\mathbf{B}}=0.
\end{eqnarray}
Using the properties that follow from expressing the fields in terms of the potentials, the constraint equations reduce to 
\begin{eqnarray}
&&\nabla\cdot\mathbf{E}-\mathbf{B}\cdot\nabla\theta  =  \rho,\\
&&\nabla\times(\mathbf{B+\theta\mathbf{E}})-\partial_{t}(\mathbf{E}-\theta\mathbf{B})= \mathbf{j}.
\end{eqnarray}
Introducing the induced charge and current $\omega$
and $\mathbf{s}$, the equations read
\begin{eqnarray}
&&\nabla\cdot\mathbf{E}  =  \rho-\nabla\theta \cdot \mathbf{B}=\rho+\omega,\\
&&\nabla\times\mathbf{B}-\partial_{t}\mathbf{E}  =  \mathbf{j}-\nabla\theta\times\mathbf{E}=\mathbf{j}+\mathbf{s}.
\end{eqnarray}

Variation of the electromagnetic terms with respect to particle positions is carried out by noticing that the density of charge and current are modified when the particle trajectories change. We have, $\delta\rho=(\nabla \rho)\cdot \delta\mathbf{x}$ as well as a term $\delta(\mathbf{A}\cdot \mathbf{j})=\mathbf{A}\cdot(\rho\delta\dot{\mathbf{x}})+(\mathbf{A}\cdot\mathbf{\dot{x}}) (\delta\mathbf{x}\cdot\nabla\rho)$. Expanding these terms, using the continuity equation for the current and eliminating divergences by integration by parts we recover the Lorentz force and we have  
\begin{equation}
\dot{\mathbf{p}}=\rho(-\nabla\phi+\dot{\mathbf{A}}+\mathbf{v}\times(\nabla\times\mathbf{A}))+\mathbf{f}_{c}
\end{equation}
with $\mathbf{f}_{c}$ the force density arising from the potential term in the matter Lagrangian. 

\section{Axion Hamiltonian}

To obtain the Hamiltonian we start with a new version of the Lagrangian
where time derivatives are replaced by velocity variables: $\mathbf{v}=\dot{\mathbf{x}}$, 
$\mathbf{V}_{E}=\dot{\mathbf{E}}$ and $\mathbf{V}_B=\dot{\mathbf{B}}$. These relations are imposed
using the multipliers $\mathbf{p}$, $\mathbf{P}_{E}$ and $\mathbf{P}_{B}$ \cite{schwinger_classical_2019mod}
\begin{eqnarray}
\mathcal{L}&=& \frac{1}{2}(-\mathbf{E}^{2}+\mathbf{B}^{2})+\theta\mathbf{E}\cdot\mathbf{B}+\frac{m}{2}\mathbf{v}^{2}\nonumber\\
 &  & \mathbf{+A}\cdot\big[\mathbf{j+V_{E}}-\theta\mathbf{V}_{B}-\nabla\times(\mathbf{B}+\theta\mathbf{E})\big]\nonumber\\
 && -\phi\big[\nabla\cdot(\mathbf{E}-\theta\mathbf{B})-\rho\big]-\mathbf{p}\cdot(\mathbf{v}-\dot{\mathbf{x}})
\nonumber \\ && -\mathbf{P_{E}\cdot(\mathbf{V_{E}-\dot{E})}-\mathbf{P_{B}\cdot(\mathbf{V_{B}-\dot{B})}}}\nonumber\\
 &  & +\mathcal{L}_{C}.
\end{eqnarray}

The equations imposed by the velocities are
\begin{eqnarray}
\mathbf{A}=\mathbf{P_{E}},\\
\theta\mathbf{A}=-\mathbf{P_{B}},\\
\rho\mathbf{A+}m\mathbf{v}=\mathbf{p}.
\end{eqnarray}
We can then write 
\begin{eqnarray}
\mathbf{A}\cdot\mathbf{j}+\frac{m}{2}\mathbf{v}^{2}-\mathbf{p}\cdot\mathbf{v} & = & 
 -\frac{1}{2m}(\mathbf{p}-q\mathbf{P_{E}})^{2}.
\end{eqnarray}

To proceed and be able to follow the typical construction of the Hamiltonian we do not eliminate, for the moment, the magnetic rate of change $\mathbf{v}_B$. Its elimination and the exclusion of the magnetic momentum is discussed as an alternative below. Elimination of the electric rate change and magnetic potential gives
\begin{eqnarray}
\mathcal{L} & = & \frac{1}{2}(-\mathbf{E}^{2}+\mathbf{B}^{2})+\theta\mathbf{E}\cdot\mathbf{B}-\frac{1}{2m}(\mathbf{p}-q\mathbf{P}_{E})^{2} \nonumber\\
 &  & -\phi\big[\nabla\cdot(\mathbf{E}-\theta\mathbf{B})-\rho\big]-\mathbf{P_{E}\cdot}\big[\nabla\times(\mathbf{B}+\theta\mathbf{E})\big] \nonumber\\
 &  & +\mathbf{P}_{E}\cdot\dot{\mathbf{E}}+\mathbf{P}_{B}\cdot\dot{\mathbf{B}}+\mathbf{p}\dot{\mathbf{x}} \nonumber\\
 &  & +\mathbf{v}_B\cdot(\theta\mathbf{P}_E-\mathbf{P}_B)+\mathcal{L}_{C}.
\end{eqnarray}
It can be seen that the retained magnetic rate change acts as multiplier connecting the momenta. 

The Hamiltonian is 
\begin{equation}
\mathcal{H}=\sum_i P_{i}\dot{X}_{i}-\mathcal{L}.
\end{equation}
Explicitly,
\begin{eqnarray}
\mathcal{H} & = & \frac{1}{2}(\mathbf{E}^{2}-\mathbf{B}^{2})-\theta\mathbf{E}\cdot\mathbf{B}+\phi\big[\nabla\cdot(\mathbf{E}-\theta\mathbf{B})-\rho\big] \nonumber\\
 &  & \mathbf{+P_{E}\cdot} \big[\nabla\times(\mathbf{B}+\theta\mathbf{E})\big]+\frac{1}{2m}(\mathbf{p}-q\mathbf{P_{E}})^{2} \nonumber\\
 &  & -\mathbf{v}_B\cdot(\theta\mathbf{P_{E}}-\mathbf{P_{B}})-\mathcal{L}_{C}.
\end{eqnarray}
The Hamiltonian equations for the magnetic field give
\begin{equation}
\dot{\mathbf{B}}=\partial_{P_{B}}H=\mathbf{v}_B
\end{equation}
and 
\begin{equation}
\dot{\mathbf{P}}_{B}=-\partial_{B}H=-\mathbf{B}+\theta\mathbf{E}+\nabla\times\mathbf{A}-\theta\nabla\phi.
\end{equation}
The equations for the electric field, $\dot{\mathbf{E}}=\partial_{P_{E}}H$, give
\begin{eqnarray}
\dot{\mathbf{E}} & = & \frac{q}{m}(\mathbf{p}-q\mathbf{P_{E}})-\nabla\times(\mathbf{B}+\theta\mathbf{E})-\theta\mathbf{v}_B\nonumber\\
 &  & =\mathbf{j}-\nabla\times(\mathbf{B}+\theta\mathbf{E})-\theta\dot{\mathbf{B}}.
\end{eqnarray}
This is the modified Ampere's equation. The momentum equation is 
\begin{equation}
{\mathbf{P_{E}}}=\dot{\mathbf{A}}=-\partial_{E}H=\mathbf{E}+\theta\mathbf{B}+\nabla\phi+\theta\nabla\times\mathbf{A}.
\end{equation}

The equations associated to particle positions give 
\begin{eqnarray}
&\dot{\mathbf{x}}&=\mathbf{j}/\rho=\partial_{p}H=\mathbf{p}-\rho\mathbf{A},
\end{eqnarray}
and
\begin{eqnarray}
 \
\dot{\mathbf{p}}& = & m\big[\dot{\mathbf{v}}+\rho\dot{\mathbf{A}}-\rho[\mathbf{v}\times\mathbf{(\nabla\times}\mathbf{A})+\mathbf{v}\nabla\cdot\mathbf{A}]\big]\nonumber \\
 &=&-q\nabla\phi+\mathbf{f}_c.
\end{eqnarray}
Here, the first equality identifies the time derivative of the momentum of a particle $\mathbf{p}=\rho\mathbf{A}+m\mathbf{v}$, which includes terms associated to the changing observed fields at its moving location. The second equality is the derivative of the field coupled to charges with respect to the charge position. 
The condition $\nabla\cdot\mathbf{A}=0$ is satisfied due to the previous
equations as well as the relation $\nabla\times\mathbf{A}=\mathbf{B}$.
Using these relations, the momentum equation becomes the Lorentz force equation. 

The potential still acts as a constraint and establishes
\begin{equation}
\nabla\cdot(\mathbf{E}-\theta\mathbf{B})=\rho.
\end{equation}

It can be checked that the equations obtained from treating the magnetic field as a canonical variable do not add new information. Therefore, it is possible to  ignore the magnetic momentum and consider the magnetic field, instead, in the same footing as the auxiliary variables $\phi$ and $\mathbf{A}$ (before it being identified with the momentum). That is, we simply exclude the equation for $\dot{\mathbf{P}}_B$ as it does not provide any new information and assume that the relation is $\mathbf{P}_B=\theta\mathbf{P}_E$ is always satisfied. This point of view leads to the expression we use at the beginning of the article. Namely, we simply consider the density $\mathcal{H}(\mathbf{E},\mathbf{B},\mathbf{A},\phi)$:
\begin{eqnarray}
\mathcal{H} & = & \frac{1}{2}(\mathbf{E}^{2}-\mathbf{B}^{2})+\theta \mathbf{E}\cdot\mathbf{B}-\phi \big[\nabla\cdot(\mathbf{E}+\theta\mathbf{B})-\rho\big] \nonumber\\  
 &  & +\mathbf{A\cdot}\nabla\times \big(\mathbf{B}-\theta\mathbf{E})+\frac{1}{2m}(\mathbf{p}-\rho\mathbf{A}\big)^2.
\end{eqnarray} 
We write $\mathbf{A}$ for the electric momentum but remembering its role as a canonical variable conjugate to the electric field and treating the magnetic field as an auxiliary variable instead of a canonical Hamiltonian variable.

Energy evaluation can be recovered at equilibrium when the constraints
are satisfied and contributes zero to the value of the Hamiltonian.
Also, $\mathbf{-P_{E}\cdot}\big[\nabla\times(\mathbf{B}+\theta\mathbf{E})\big]$
can be integrated by parts to obtain ($\nabla\times\mathbf{P_{E}})\cdot(\mathbf{B}+\theta\mathbf{E})=\mathbf{B}\cdot(\mathbf{B}+\theta\mathbf{E}).$
The final energy evaluation is 
\begin{equation}
\mathcal{E}=\frac{1}{2}(\mathbf{E}^{2}+\mathbf{B}^{2})+\frac{m}{2}v^{2}.
\end{equation}
This is, however, not a variational functional nor a Hamiltonian.  It simply indicates
the value of the energy in a physical configuration.


\bibliography{references}    
\end{document}